\documentstyle[12pt]{article}
\begin{document}
hep-th/9511135

\begin{center}
{\Large\bf On the regularization scheme and gauge choice ambiguities in
topologically massive gauge theories}
\medskip
\medskip

{J. Gegelia, A. Khelashvili and
N. Kiknadze}
\medskip

{\it Department of General Physics and High Energy Physics
Institute, Tbilisi State University,
Chavchavadze ave.\ 1, 380028 Tbilisi, Republic of Georgia}

\medskip

\begin{abstract}
It is demonstrated that in the $(2+1)$-dimensional topologically massive
gauge theories an agreement of the Pauli-Villars
regularization scheme with the other schemes can be achieved by
employing pairs of
auxiliary fermions with the opposite sign masses. This approach does not
introduce additional violation of discrete ($P$ and $T$) symmetries.
Although it breaks the local gauge symmetry only in the regulator fields'
sector, its trace disappears completely after removing the
regularization as a result of superrenormalizability
of the model. It is shown also that analogous extension of the
Pauli-Villars regularization in the vector particle sector can be used to
agree the arbitrary covariant gauge results with the Landau ones. The
source of ambiguities in the covariant gauges is studied in detail. It is
demonstrated that in gauges that are softer in the infrared region
(e.g.\ Coulomb or axial) nonphysical ambiguities inherent to the covariant
gauges do not arise.
\end{abstract}

{ PACS numbers 11.10.Kk, 11.15.-q, 11.10.Gh}
\end{center}
\medskip

\section{Introduction}

Since the extensive work after Deser, Jackiw and Templeton \cite{1} it is
well known that in the $(2+1)$-dimensional  topologically massive
gauge theories there arise several ambiguities at the one loop
level. First of all we mean the regularization scheme and gauge
parameter dependence of physical quantities and some other,
related to these, problems. The most distinct among them are the
regularization scheme ambiguities. The problem is that at one loop
approximation the Pauli-Villars (PV) method in contrast to the other
regularization schemes produces different gauge field polarization
operators \cite{2}--\cite{8}. Another problem is that in calculation of
the fermion self-energy the Landau gauge plays an outstanding role
among the other covariant gauges \cite{1}. The fermion pole-mass
in an arbitrary covariant gauge turns out to be gauge parameter dependent.
This trouble can be avoided if we use the auxiliary vector fields as PV
regulators, but even in this case
the numerical result differs from that of the Landau gauge \cite{1}.

In $(2+1)$ dimensions the PV regularization scheme unlike
dimensional (or the other) regularization schemes introduces parity
violation at the intermediate stage, the finite trace of which
survives after removing PV regularization. In our previous publication
\cite{9} we have demonstrated that in abelian case it is possible to agree
polarization operator calculations in all schemes by using pairs of
opposite sign mass fermions as regulators. Below we are going to show
that this result can be generalized for the nonabelian case too.
We are going to study also the consistency problem of different
gauges with
the help of pairs of regulator vector fields i.e.\ we will show that
the fermion pole mass calculated in an arbitrary covariant gauge
coincides with  the Landau gauge result if the gauge parameters for
the pairs of regulator vector fields with opposite sign masses
are set the same as that of the initial gauge field.
As long as the gauge propagator in the covariant gauge is more infrared (IR)
singular than in the other ones, we will study the same problem in the IR
softer gauges (Coulomb and axial). we will show that they produce
the Landau gauge result. Moreover it turns out that the nonphysical cut
present in the Landau gauge disappears in IR softer gauges.

\section{Polarization operator of the gauge field and the
three-gluon vertex at one loop}

Consider QED (or QCD) with massive fermions plus Chern-Simons term
in $(2+1)$-dimensions. Below we will follow notations from
\cite{1}. Due to the superrenormalizability of these models only
few one-loop diagrams are divergent. In this section we study the
gauge field polarization operator:
\begin{equation}
\Pi_{\mu\nu}(k,m)=-ig^2\int{\frac{d^3q}{(2\pi)^3}tr\left[\gamma_\mu
S(k+q)\gamma_\nu S(q)\right]}
 \label{2.1}
\end{equation}
and the three-gluon vertex:
\begin{equation}
\Pi_{\mu\nu\alpha}(k,q,m)=-g^3\int{\frac{d^3p}{(2\pi)^3}
tr\left[\gamma_\mu
S(p)\gamma_\nu S(p+k)\gamma_\alpha S(p+k+q)\right]}\ .
              \label{2.2}
\end{equation}
Here $S(q)=-i(\hat q-m)^{-1}$. Explicit calculation with the cutoff
$\Lambda$ yields \cite{1}, \cite{5}:
\begin{equation}
\Pi_{\mu\nu}(k,m)=-\frac{g^2}{3\pi}\Lambda g_{\mu\nu}
+\left(g_{\mu\nu}-\frac{k_\mu k_\nu}{k^2}\right)\Pi^{(1)}(k^2,m)+
i\epsilon_{\mu\nu\lambda}k^{\lambda}\Pi^{(2)}(k^2,m)
              \label{2.3}
\end{equation}
with
\begin{equation}
\Pi^{(1)}(k^2,m) =
\frac{g^2}{2\pi}\left[\frac{\sqrt{m^2}}{2}
-\left(\frac{m^2}{2}+\frac{k^2}{8}\right)\frac{1}{\sqrt{k^2}}
ln\left(\frac{2\sqrt{m^2}+\sqrt{k^2}}
{2\sqrt{m^2}-\sqrt{k^2}}\right)\right]\ ,
\label{2.4a}
\end{equation}
\begin{equation}
\Pi^{(2)}(k^2,m)=
\frac{g^2}{4\pi}
\frac{m}{\sqrt{k^2}}
ln\left(\frac{2\sqrt{m^2}+\sqrt{k^2}}
{2\sqrt{m^2}-\sqrt{k^2}}\right)\ .
\label{2.4b}
\end{equation}
If we use the dimensional regularization result will turn out finite
and (\ref{2.3}) without the first term and (\ref{2.4a})--(\ref{2.4b})
will be final expressions. In the PV scheme we have to consider
combination:
\begin{equation}
\Pi_{\mu\nu}^{Reg}(k)=\sum_{i=1}^{N}C_i\Pi_{\mu\nu}(k,M_i)
\end{equation}
where $C_1=1$, $M_1=m$ and the limit $\mid M_{i>1}\mid\to\infty$ is
to be taken. From the above expressions it is clear that
\begin{equation}
\Pi^{(1)}(k^2,\mid M_i\mid\to\infty) = -\frac{g^2}{12\pi}
\frac{k^2}{\sqrt{M_i^2}}\to 0\ ,   \label{2.6}
\end{equation}
\begin{equation}
\Pi^{(2)}(k^2,\mid M_i\mid\to\infty) = \frac{g^2}{4\pi}
\frac{M_i}{\sqrt{M_i^2}} = \frac{g^2}{4\pi} sgn(M_i)\ . \label{2.7}
\end{equation}
If only one regulator field is used, we obtain:
\begin{equation}
\Pi^{(1)}_{PV}(k^2,m)=\Pi^{(1)}(k^2,m)\ ,
\label{2.8}
\end{equation}
\begin{equation}
\Pi^{(2)}_{PV}(k^2,m)=\Pi^{(1)}(k^2,m)-\frac{g^2}{4\pi} sgn(M)\ .
\label{2.9}
\end{equation}
For massless fermions we have:
\begin{eqnarray}
\Pi^{(1)}_{PV}(k^2,m=0) & = & \frac{g^2\sqrt{-k^2}}{16}\ , \nonumber \\
\Pi^{(2)}_{PV}(k^2,m=0) & = & -\frac{g^2}{4\pi} sgn(M)\ . \label{2.10}
\end{eqnarray}
While the dimensional regularization yields:
\begin{eqnarray}
\Pi^{(1)}_{Dim}(k^2,m=0) & = & \frac{g^2\sqrt{-k^2}}{16}\ , \nonumber \\
\Pi^{(2)}_{Dim}(k^2,m=0) & = & 0\ .  \label{2.11}
\end{eqnarray}
So radiative corrections to the Chern-Simons term for massless
fermions are absent in the dimensional regularization while in the
PV one they are nonzero and depend on the sign of the regulator fermion
mass!
The latter is evidently absurd result --- for massless theory there
is no source of contribution to the antisymmetric
$\epsilon_{\mu\nu\lambda}$ structure and its
appearance is an artefact of the PV regularization itself. Of course
in the other regularization schemes, which respect discrete ($P$, $T$)
symmetries at the intermediate stage, no such contribution arises
\cite{5}. Hence it is natural to modify the PV method in a way to
cancel the radiative corrections to the antisymmetric structure. It
can be achieved if, apart from the condition of cancellation of
divergence
\begin{equation}
1+\sum_{i=2}^{N}C_i=0\ ,          \label{2.12}
\end{equation}
another condition is also demanded \cite{9}:
\begin{equation}
\sum_{i=2}^{N}C_i sgn(M_i)=0\ .  \label{2.13}
\end{equation}
Note first of all that it is impossible to satisfy both (\ref{2.12}) and
(\ref{2.13}) if all regulators have the same sign masses. It is
evident also that some of the coefficients $C_i$ have to be
fractional. E.g.\ for two regulator fermions with opposite sign
masses solution of (\ref{2.12}) and (\ref{2.13}) is
\begin{equation}
C_2=C_3=-\frac{1}{2}\ .        \label{2.14}
\end{equation}
Fractional coefficients make difficult conventional counterterm
interpretation of the PV procedure \cite{10}. But alternatively
we can interpret fractional coefficients as being originated from
fractional charges of the regulator fermions, i.e., their coupling
constants to the gauge fields are $\mid C_i\mid^{\frac{1}{2}}g$ instead
of $g$. This interpretation is consistent for the abelian theory
\cite{11}, but for the nonabelian one it causes local gauge symmetry
violation in the auxiliary fields' sector --- in counterterms. The question
is whether it will affect the final (after removing regularization)
results. To find answer let us examine the nonabelian case which, as
a rule, can make things clear. Here the
PV regularization modifies the three-gluon vertex (\ref{2.2}) too and
the extra term is again proportional to the sign of the mass of the
regulator fermion \cite{6}:
\begin{equation}
\lim_{\mid M\mid\to\infty}\Pi_{\mu\nu\alpha}(p,k;M)=
-\epsilon_{\mu\nu\alpha}\frac{g^3}{4\pi} sgn(M)\ .  \label{2.15}
\end{equation}
Hence using the counterterms described above we will get an extra
term:
\begin{equation}
g^3 \sum_{i=2}^{N}\mid C_i\mid^{\frac{3}{2}} sgn(M_i)\ . \label{2.16}
\end{equation}
In general, it is not clear whether this term will vanish provided
(\ref{2.12}) and (\ref{2.13}) are satisfied except when all of the
coefficients $C_i$ are the same: $C_i=-1/(N-1), N\geq 2$.
Then (\ref{2.13}) takes the following form:
\begin{equation}
\sum_{i=2}^{N} sgn(M_i)=0            \label{2.17}
\end{equation}
and (\ref{2.16}) vanishes if (\ref{2.17}) is satisfied. As for the
higher orders or other Green's functions, the terms generated by
counterterms vanish after removing regularization and hence leave no
trace in the final results.

Condition (\ref{2.17}) is quite interesting from the physical
point of view. Evidently, it can be satisfied only if we take equal
number of the regulator fermions with both signs of masses. It means
that in the limit of the infinite regulator masses  the regulator
Lagrangian would preserve parity. Of course it does not matter how many
pairs with opposite mass signs are taken. Therefore more than one pair of
the regulators is abundant.

Thus if we use the above described generalized PV scheme which preserves
parity, it will produce results identical to the other parity preserving
schemes
It seems quite natural for us because the only distinction of principle
of the ordinary PV scheme with the odd number of fermions is its
parity-violating nature.
Note that in the massive model parity is not a symmetry of the
theory. But as in many similar cases, the counterterms of a
parity-preserving (symmetric) model are sufficient to renormalize
a parity-violating (broken symmetry) massive model as well.

\setcounter{equation}{0}
\section{The fermion propagator}

We will see below that in the case of the fermion propagator one
faces the gauge choice problem.
In an arbitrary covariant gauge the gauge field propagator has the form
\cite{1}:
\begin{equation}
D_{\mu\nu}=\frac{-i}{p^2-\mu^2+i0}(g_{\mu\nu}-\frac{p_\mu p_\nu}{p^2}
-\frac{i\mu}{p^2}\epsilon_{\mu\nu\lambda}p^\lambda)-i\xi\frac{p_\mu
p_\nu}{p^4}         \label{3.1}
\end{equation}
where $\xi$ is the gauge parameter and $\mu$ is the Chern-Simons
constant. The corresponding gauge fixing Lagrangian is
\begin{equation}
{\cal L}_{gf}=-\frac{1}{2\xi}\left(\partial_\mu A^\mu\right)^2\ .
         \label{3.2}
\end{equation}
Below we will use the one-loop results from \cite{1}. The fermion
self-energy operator is made up from three terms
\begin{eqnarray}
\Sigma_I(p) & = & \frac{-g^2}{16\pi}\int_{-\infty}^{\infty}
\frac{da}{\hat
p-a}\biggl[\left(\frac{\mu^2}{a^2}+\frac{4m}{a}\right)
\theta(a^2-{\cal M}^2) + \nonumber \\
 & + & \frac{1}{\mu^2a^2}(a^2-m^2)^2\theta({\cal M}^2-a^2)
\theta(a^2-m^2)\biggr]\ ,
              \label{3.3a}
\end{eqnarray}
\begin{eqnarray}
\Sigma_{II}(p) & = & \frac{-g^2}{8\pi}\int_{-\infty}^{\infty}
\frac{da}{\hat
p-a}\biggl[\left(a+m\right)\frac{\mu}{a^2}\theta(a^2-{\cal M}^2)
+ \nonumber \\
 & + & \frac{(a-m)(a^2-m^2)}{\mu a^2}\theta({\cal M}^2-a^2)
\theta(a^2-m^2)\biggr]\ ,
              \label{3.3b}
\end{eqnarray}
\begin{equation}
\Sigma_{III}(p)  =  \frac{-g^2}{16\pi}\int_{-\infty}^{\infty}
\frac{da}{\hat p-a}\frac{(a+m)^2}{a^2}\theta(a^2-m^2)\ ,
              \label{3.3c}
\end{equation}
$$
{\cal M}\equiv m+\mid\mu\mid\ .
$$
Here contributions of the antisymmetric ($\Sigma_{II}$) and the gauge
fixing terms ($\Sigma_{III}$) are separated. Expressions
(\ref{3.3a})--(\ref{3.3c})
can be obtained by arbitrary Lorentz invariant integration and hence by
the dimensional regularization too. Result is finite --- expected
logarithmic divergence has disappeared. As for the regulator
fermions in the PV regularization mentioned in the previous section, they
do not contribute here. Although the IR singularity of the massless
longitudinal gauge particles is integrable, there arise some pathologies
in the mass shell behaviour \cite{1}. Namely a)~$\Sigma_{III}(m)\neq
0$ and hence the fermion renormalized (pole) mass is gauge parameter
($\xi$) dependent, b) $\Sigma_{III}'(m)$ diverges and the wave
function renormalization can not be defined.

These pathologies can be avoided in the Landau ($\xi=0$) gauge
where  $\Sigma_{III}$ is absent. Hence the Landau gauge plays
distinguished role.

Can the above mentioned troubles be avoided for $\xi\neq 0$?
For that purpose authors of \cite{1} considered the PV
regularization using regulator vector fields. Sources of troubles
reside in the infrared region and it may seem strange that the PV
regularization
cures them. But $\Sigma_{III}(p)$ does not depend on the vector particle
mass. Therefore its contribution to the total fermion self energy will
cancel in the PV scheme if we take the gauge parameter of the regulator
vector partricle to be equal to the original one. Unfortunately this is not
the end of the story --- after removing regularization there will
arise additional contribution to the $\Sigma_{II}$ structure. Indeed
\begin{equation}
\left.\Sigma_{II}(p)\right\vert_{\mid\tilde\mu\mid\to\infty} =
-\frac{g^2}{2\pi}
\frac{\tilde\mu}{\mid\tilde\mu\mid}=-\frac{g^2}{2\pi}sgn(\tilde\mu)\ .
\label{3.4}
\end{equation}
Here $\tilde\mu$ denotes the mass of the regulator vector particle.
So the PV regularization result (e.g.\ for pole-mass) will differ from
dimensionally regularized Landau gauge result due to the
additional contribution (\ref{3.4}).

The way out can be found if we notice that the additional term
(\ref{3.4}) is proportional to the sign of regulator mass and
therefore we can extend the PV procedure in full analogy to the previous
section. Namely, we can introduce several regulator vector
fields and require that the following conditions are satisfied:
\begin{equation}
1+\sum_{i=2}^{N}C_{i}'=0\ , \qquad\qquad \sum_{i=2}^{N}
C_i'sgn(\tilde\mu_i)=0\ .
\label{3.5}
\end{equation}
The meaning of the coefficients $C_i'$ is the same as in the previous
section and $\tilde\mu_i$ are the masses of the regulator gauge
fields. Again, the fact that these auxiliary gauge fields interact
with matter fields (and in nonabelian case selfinteract too) with
couplings $g\mid C_i\mid^\frac{1}{2}$ does not violate the gauge
invariance of final expressions. Apparently if the gauge parameters
for these fields were chosen to be equal then any covariant gauge
would reproduce the dimensionally regularized Landau gauge results.

It is worth noting that quite similarly to the conditions
(\ref{2.12})--(\ref{2.13}) conditions (\ref{3.5}) have solutions that
correspond to the parity preservation in the end --- pairs of regulators
with opposite sign masses. The simplest choice is just one pair with
\begin{equation}
C_2'=C_3'=-\frac{1}{2}\ .    \label{3.6}
\end{equation}

So, described generalization of the PV scheme leads to the agreement
of arbitrary covariant gauges. But evidently, introduction of any
additional procedure for agreeing the covariant gauges seems to be quite
artificial and has no well argumented theoretical basis.

\setcounter{equation}{0}
\section{The fermion propagator in the axial gauge}

{}From the discussion in the previous section it is clear that the Landau
($\xi=0$) gauge is the only applicable one out of the covariant gauges
because it is free of the infrared singularities. Still, even in the
Landau gauge, there remains the following problem: after calculating
the pole mass and defining the wave function renormalization
constant $Z_F$, one finds that there is a nonphysical cut in the
continuum contribution starting at $p^2=m^2$, i.e.\ earlier than
the two-particle threshold at $p^2=(m+\mu)^2$ \cite{1}. The reason
is the survival of the contributions of virtual, nonphysical processes
originated by $p_\mu p_\nu/p^2$ and $\epsilon_{\mu\nu\alpha}p_\alpha/p^2$
terms of the propagator in the Landau gauge.

The question whether this superfluous contribution does not show up in the
physical processes (e.g.\ the Compton scattering amplitude etc.) is not
{\it a priori} clear. Below we will see that this problem is absent
in the IR softer gauges.

First let us consider an arbitrary axial gauge, when
${\cal L}_{gf}=-\frac{1}{2\xi}(nA)^2$ with $n_\mu$ being some
spacelike vector, $n^2<0$. The propagator of the vector field has
the form \cite{13}:
\begin{equation}
D_{\mu\nu}=\frac{-i}{p^2-\mu^2}(g_{\mu\nu}-\frac{p_\mu n_\nu+p_\nu
n_\mu}{np}+\frac{n^2 p_\mu p_\nu}{(np)^2}
-\frac{i\mu}{np}\epsilon_{\mu\nu\lambda}n^\lambda)-i\xi\frac{n^2 p_\mu
p_\nu}{(np)^2}\ .                \label{4.1}
\end{equation}

If $\xi=0$ the general axial gauge is reduced to the homogeneous $nA=0$
gauge \cite{14}. The $\xi$-dependent part does not contribute to the
fermion pole mass because the corresponding integral is less
singular compared to the covariant gauge. Indeed, in the axial gauge
we have:
\begin{equation}
\delta m_\xi\sim g^2\left.\xi\frac{\hat p}{p^2}
(p^2-m^2)^2
\int d^3q\frac{1}{(nq)^2((p-q)^2-m^2)}\right\vert_{\hat p=m}=0\ ,
              \label{4.2}
\end{equation}
while in the covariant gauge we have:
\begin{equation}
\delta m_\xi\sim g^2\left.\xi\frac{\hat p}{p^2}
(p^2-m^2)^2
\int d^3q\frac{1}{q^4((p-q)^2-m^2)}\right\vert_{\hat p=m}=2i\pi^2g^2\xi\ .
\end{equation}

Hence we can neglect the $\xi$-dependent term and check the gauge
$n_\mu$-vector independence of physical quantities. At this stage
there arise two problems inherent in axial gauge. The first one is
the problem of prescription for singular $(np)^{-k}$ denominators. It
was demonstrated in \cite{15} that it is possible to use any
prescription ([np]) that in the sense of generalized functions
\cite{16} satisfies condition $np/[np]=1$. Another problem
is the form of the correct equation for the fermion pole mass. The
usual equation
\begin{equation}
\left.S^{-1}_F(p)U(p)\right\vert_{p^2=m^2}=0   \label{4.3}
\end{equation}
(with $S_F$ being the fermion propagator) is useless in the axial
gauge except for the choice $n_\mu\sim p_\mu$. Quite often the
following equation is used \cite{17}:
\begin{equation}
\left.\bar U(p)S_F^{-1}(p)\right\vert_{\hat p=m}U(p) = 0\ .
              \label{4.4}
\end{equation}
In \cite{18} it was demonstrated that this equation produces correct
results only at one loop. Beginning from the second loop its
solution becomes $n_\mu$-vector dependent. The correct approach is to
equate the denominator of the fermion propagator to zero only
after rationalizing it (i.e.\ after eliminating the $\gamma$-matrices
from the denominator). Anyway our analysis does not exceed one loop and
we can use (\ref{4.4}) as well.

The (\ref{4.1}) propagator (with $\xi$ set to zero) results the
following contributions to the fermion self energy operator:
\begin{eqnarray}
\Sigma_{g_{\mu\nu}}(p) & = & -\frac{ig^2}{(2\pi)^3}\int d^3q\biggl\{
\frac{3m-\hat p}{\left(q^2-\mu^2\right)
\left(\left(p^2-q^2\right)-m^2\right)}+\frac{\hat p}{2p^2}
\frac{m^2-\mu^2}{\left(q^2-\mu^2\right)\left(q^2-m^2\right)}+
\nonumber \\
& + & \frac{\hat p}{2p^2}\frac{p^2-m^2+\mu^2}{\left(q^2-\mu^2\right)
\left(\left(p^2-q^2\right)-m^2\right)}\biggr\}\ ,  \label{4.5}
\end{eqnarray}
\begin{eqnarray}
\Sigma_{(np)^{-1}}(p) & = & -\frac{ig^2}{(2\pi)^3}\int d^3q\biggl\{
\frac{\left(m^2-p^2\right)\hat n}{(nq)\left(q^2-\mu^2\right)
\left(\left(p^2-q^2\right)-m^2\right)}+ \nonumber \\
 & + & \frac{\left(\hat p-m\right)\hat q\hat n+\hat n\hat q
 \left(\hat p-m\right)}{(nq)\left(q^2-\mu^2\right)
\left(\left(p^2-q^2\right)-m^2\right)}\biggr\}\ ,
                                             \label{4.6}
\end{eqnarray}
\begin{eqnarray}
\Sigma_{(np)^{-2}}(p) & = & -\frac{ig^2n^2}{(2\pi)^3}\int d^3q\biggl\{
\frac{m-\hat p}{\left(q^2-\mu^2\right)(nq)^2}+
\frac{(\hat p-m)(p^2-m^2)}{(nq)^2\left(q^2-\mu^2\right)
\left(\left(p^2-q^2\right)-m^2\right)}+ \nonumber \\
 & + &
\frac{(\hat p-m)\hat q(\hat p-m)}{(nq)^2\left(q^2-\mu^2\right)
\left(\left(p^2-q^2\right)-m^2\right)}\biggr\}\ ,
                                         \label{4.7}
\end{eqnarray}
\begin{eqnarray}
\Sigma_{\mu}(p) & = & -\frac{i2\mu g^2}{(2\pi)^3}\int d^3q\biggl\{
\frac{1}{\left(q^2-\mu^2\right)
\left(\left(p^2-q^2\right)-m^2\right)}+ \nonumber \\
&+&\frac{(np)-m\hat n}{(nq)\left(q^2-\mu^2\right)
\left(\left(p^2-q^2\right)-m^2\right)}\biggr\}\ .  \label{4.8}
\end{eqnarray}
For definiteness we choose $m>0$. Examining expressions
(\ref{4.5})--(\ref{4.8}) it is easy to see that the second terms in
$\Sigma_\mu(p)$ and $\Sigma_{(np)^{-1}}(p)$ will cause problems due
to the $\hat n$ dependence if one tries to employ the standard equation
(\ref{4.3}). At the same time it is clear that in (\ref{4.4}) these
terms do not contribute and the nonvanishing contributions
(resulting in by $\Sigma_{g_{\mu\nu}}$ and the first term in $\Sigma_{\mu}$)
are independent of $n_\mu$. Explicit calculation of the pole mass
using dimensional regularization results:
\begin{equation}
\delta m =
 \frac{g^2}{16\pi}\biggl[(2-\frac{\mu}{m})^2ln(1+2\frac{m}{\mid \mu\mid})
+ 2(1-\frac{\mid \mu\mid}{m})\biggr]\ .     \label{4.9}
\end{equation}

Note that in the $\mid \mu\mid\to 0$ limit this expression diverges. So
the topological (originated by CS term) mass plays the role of
infrared regulator. On the other hand if we use the PV
regularization introducing a single auxiliary vector field, then in
place of (\ref{4.9}) we have:
\begin{equation}
\left.\delta m\right\vert_{PV} = \left.\delta m\right\vert_{dim} -
\left.\delta m\right\vert_{dim,\mid \mu_{reg}\mid\to\infty} =
\left.\delta m\right\vert_{dim} + \frac{g^2}{2\pi}sign(\mu_{reg})\ .
              \label{4.10}
\end{equation}
It coincides with (\ref{3.4}), i.e.\ there is a similar discrepancy between
dimensional and PV regularizations in the axial gauge too. Hence
this discrepancy can be avoided in the same manner by introduction
of pairs of opposite mass sign regulator vector particles.

Evidently
$\left.\delta m\right\vert_{m\to 0}=-\frac{g^2}{2\pi}sgn(\mu)$,
i.e.,
even if we start from massless theory, then due to the CS term the
fermion mass will be generated. But the vector $PV$ regularization will
predict different values for the fermion mass. Moreover, for the specific
choice of number and mass
signs of auxiliary vectors it is possible to avoid generation of the mass.

Note also that when $n_\mu$ is a constant vector, the
$n_\mu$-dependent terms of the gauge field propagator do not
contribute to the physical quantities, while the $n_\mu$ independent
ones have the structure:
$$
\int \frac{d^3q}{\left(q^2-\mu^2\right)
\left(\left(p-q\right)^2-m^2\right)}\ .
$$
These terms will produce physical cut beginning at $p^2=(m+\mu)^2$.
So the nonphysical cut (starting from $p^2\geq m^2$) present in the
Landau gauge is absent in the axial gauge.

At the end let us mention that we have calculated the fermion
pole mass also in the ``most physical" Coulomb gauge. We have
made use of the propagator
\begin{equation}
D_{\mu\nu}=\frac{-i}{p^2-\mu^2+i0}(g_{\mu\nu}+
\frac{\bar p_\mu\bar p_\nu}{{\vec p}{\,}^2}-
\frac{n_\mu n_\nu p_0^2}{{\vec p}{\,}^2}+\frac{i\mu}{{\vec p}{\,}^2}
\epsilon_{\mu\nu\lambda}\bar p^\lambda)\ ,
              \label{4.11}
\end{equation}
$$
\bar p^\mu=(0,\vec p)\ , \qquad\qquad n^\mu=(1,\vec 0)\ ,
$$
Its time-time component is not integrable in two-dimensional space.
In \cite{1} it was admitted that the infrared safe version of
(\ref{4.11}) (i.e.\ without the second and third terms) produces the
pole mass identical to the Landau gauge. We have checked by explicit
calculation that the same result can be obtained using (\ref{4.11}).
Moreover, like the axial gauge, the nonphysical cut is absent in the
Coulomb gauge too.

\section{Conclusions}

Above we have considered ways and means to avoid undesirable
nonphysical ambiguities arising in $(2+1)$-dimensional topologically
massive gauge theories. The above described generalization of the
Pauli-Villars regularization seems to us to be the most interesting.
Quite nontrivial feature is that the local gauge symmetry violated
by the counterterm sector is restored after removing
regularization. The crucial role is played by superrenormalizability
of the models under consideration --- due to this feature
only finite number of the one loop diagrams are affected by the
regularization. The ambiguities in the ordinary PV regularization are just
artefacts of the regularization itself and it seems quite natural to
avoid these ambiguities by preserving discrete symmetries in the
auxiliary sector. It is evident that the described PV regularization scheme
excludes, for example, appearence of the parity anomaly \cite{19} and
of the other analogous effects as well. Another
question is why should we employ the modified PV regularization,
i.e., why should we specify the regularization scheme together with
quantization of the model? There is no answer to this question in general.
We can only require that the regularization used should not produce
physically meaningless results. The suggested modification of the PV
scheme satisfies this requirement and is compatible with the other
schemes which does not introduce additional violation of discrete
symmetries. The same is true about the existence of preferred gauge,
such as the Landau gauge out of covariant gauges though even in the
Landau gauge there remains trace of nonphysical processes resulting
nonphysical cut in the fermion propagator. We have demonstrated
above that by the means of generalization of the PV regularization in
the vector particles' sector the compatibility of all covariant
gauges (including the Landau gauge) can be achieved. The PV
procedure in this case takes care of IR but not UV
problem. The considered example shows that one must take special
care while introducing gauge fixing terms into Langrangian
especially when gauge variant quantities like the Green's functions
are considered. As for the noncovariant gauges (e.g.\ axial or
Coulomb), they, being ``softer" in infrared, do not imply nonphysical
singularities.

\end{document}